\documentclass[aps,prl,superscriptaddress,twocolumn,floatfix]{revtex4-1}
\usepackage{graphicx}
\usepackage{dcolumn}
\usepackage{bm}
\usepackage{hyperref}
\usepackage{amssymb}
\usepackage{amsmath}
\usepackage{array}

\usepackage{rotating} 

\usepackage{siunitx}

\usepackage{natbib}
\usepackage{epstopdf}
\usepackage{color}
\usepackage{makecell}
\usepackage[table]{xcolor}
\usepackage{tabularx}

\usepackage{lmodern}
\usepackage[format=plain,justification=centerlast]{caption}
\usepackage[latin1]{inputenc} 

\graphicspath{{Figure_eps/}} 

\definecolor{lightgray}{gray}{0.9}
\definecolor{darkgray}{gray}{0.60}
\newcolumntype{C}{c<{\kern\tabcolsep}<{\kern\tabcolsep}@{}}
\begin{document}
\clearpage

\newcommand{\be}{\begin{equation}}
\newcommand{\ee}{\end{equation}}
\newcommand{\del}{\partial}

\let\oldAA\AA
\renewcommand{\AA}{\text{\normalfont\oldAA}}

\newcommand{\LL}[1]{\textcolor{red}{{\bf LL:} #1}}
\newcommand{\MT}[1]{\textcolor{red}{{\bf MT:} #1}}
\newcommand{\LH}[1]{\textcolor{blue}{{\bf LH:} #1}}
\newcommand{\JFJ}[1]{\textcolor{purple}{{\bf JFJ:} #1}}
\newcommand{\PBRO}[1]{\textcolor{cyan}{{\bf PBRO:} #1}}

\preprint{APS/123-QED}

\newcommand{\udem}{D\'{e}partement de Physique and Regroupement
 Qu\'{e}b\'{e}cois sur les Mat\'{e}riaux de Pointe, Universit\'{e} 
de Montr\'{e}al, C.P. 6128, Succursale Centre-Ville, Montr\'{e}al,
 Qu\'{e}bec, Canada H3C~3J7}
\newcommand{\nrc}{National Research Council of Canada, Ottawa, On.,
 Canada K1A 0R6}

\title{A critical assessment of models of
pair-interactions and screening used in analyzing recent warm-dense matter experiments.} 

\author{L. Harbour}\affiliation{\udem}
\author{M. W. C. Dharma-wardana}\affiliation{\nrc}
\email[Email address:\ ]{chandre.dharma-wardana@nrc-cnrc.gc.ca}
\author{D. D. Klug}\affiliation{\nrc}
\author{L. J. Lewis}\affiliation{\udem}

\date{\today}

\begin{abstract} 
Ultra-fast laser experiments yield  increasingly reliable
data on warm-dense matter (WDM), but  rely on
entrenched simplistic theoretical models. 
We re-analyze two topical  experiments,  avoiding  (i)  {\it ad hoc}
core-repulsion models,  (ii) ``Yukawa screening'' models  and (iii)
electron-ion equilibrium assumptions. An accurate, rapid
  density-functional neutral-pseudoatom model coupled to a
hyper-netted-chain (HNC) equation with a bridge term is  used to  compute
structure factors, X-Ray scattering, compressibility, phonons and
resistivity. Electronic-structure codes are used
to confirm the calculations. The Yukawa and core-repulsion models are
 shown to be misleading.
 \end{abstract}

\maketitle

{\it Introduction}$-$ 
 High-energy deposition using ultra-fast lasers has  created
novel non-equilibrium regimes of density and temperature,
 raising issues of broad scientific interest. The topics cover
hollow atoms, quasi-solids and transient plasmas.
 The physics of such  warm-dense matter (WDM) applies to
 hot-carriers in
nanostructures, space re-entry, protective shields against photonic weapons,
inertial confinement fusion ~\cite{Dimonte08,ICF-Graz11}, Coulomb explosions,
laser machining  and ablation~\cite{Lewis-Abl-03}, and in astrophysics. 
 WDM systems are strongly-correlated, with the effective coupling
parameter $\Gamma$ (ratio of the Coulomb energy to the kinetic energy) 
 greatly exceeding unity.  

WDM systems are created using, e.g., (i) shock-compression
~\cite{Flet-Al-15,Saiz-Li-08, Ma-Al-13} and (ii) ultra-fast laser heating
\cite{Milsch-88,chen2013,Sper15}. In ultra-fast heating the femto-second  optical
pulses  directly heat  the electrons, increasing their
temperature $T_e$ to many eV, while the ions remains nearly at their initial
temperature  and density $\rho^0$. Such WDM systems are  referred to
as ultra-fast matter (UFM). At  short time
 scales ( e.g., $<$100 fm) even the electrons
may  not equilibrate to  a temperature~\cite{Medvadev}. The ions and
electrons equilibrate in timescales exceeding the electron-ion relaxation
time $\tau_{ei}$, i.e., over  hundreds of picoseconds~\cite{chen2013,elr2001}.
Thus, experiments with  $t<\tau_{ei}$ deal with non-equilibrium UFM.
 The simplest non-equilibrium paradigm  is the well-known two-temperature
 (2$T$) model~\cite{Anisimov74}. Nevertheless, many  WDM-UFM studies 
have used equilibrium models, although later  shown to need at least
 a $2T$ model~\cite{cdwPlasmon,Plag15,Clerouin}.  
Electronic-structure  codes~\cite{VASP,Abinit} based on density
 functional theory (DFT) coupled with molecular dynamics (MD) are  used for
interpreting these experiments. 

The results from DFT-MD have themselves been
fitted to intermediate quantities like pair-potentials to harvest more physics
and simplify the computations. Simple intuitive models usually have hidden 
pitfalls that  become entrenched unless corrected. The objective of this study
is to re-analyze two recent
experiments~\cite{Flet-Al-15,Ma-Al-13,Saiz-Li-08} using the  DFT
 neutral-pesudo-atom(NPA)  approach which is as accurate as the DFT-MD codes
for many systems, but orders of magnitude
faster. It directly yields physically useful quantities like
pair potentials and structure factors  needed in
computing observed properties.  

Many  WDM studies during the last seven
years, e.g., in Refs.\cite{Flet-Al-15,Ma-Al-13} have used  an intuitive
 ``Yukawa + short-ranged repulsive (YSRR) potential''
 $\beta_i V_{ii}^{\text{ysrr}}(r) =
\sigma^4/r^4+\beta_i\exp(k_sr)/r$ introduced in Ref.~\cite{Wunsch09}.
$T_i = 1/\beta_i$ is the ion temperature, $k_s$ is a screening
wavevector and $\sigma$ is a parameter fitted to MD data.  We examine its validity
using first-principles models and XRTS data for ultrafast
($T_e\ne T_i$) as well as equilibrium systems. The YSRR potential is
found to produce  misleading conclusions.

XRTS data yields  $T_e$, $T_i$, ion density $\rho$,
electron density $n_e$, and details of ionic and electronic correlations.
 An important component of the XRTS signal is
the  \textit{ion feature} $W(k)$.~\cite{GlenzerRMP,Gregori06} 
\begin{eqnarray}
\label{xrts-eqn}
W(\mathbf{k},\omega) &=& |f(\mathbf{k})+ 
q(\mathbf{k})|^2 S_{ii}(\mathbf{k},\omega)\\
S_{ii}(\mathbf{k},\omega) &\simeq& S_{ii}
(\mathbf{k})\delta(\omega)
\end{eqnarray}
Here $f(\mathbf{k})$ and $q(\mathbf{k})$ are the form factors of bound and
 free electron densities at an ion.  $S_{ii}(\mathbf{k},\omega)$ is the
 dynamic structure factor of the ions. Current
 XRTS cannot  resolve ion dynamics (at meV energy scales).
 It  is  approximated via the static
 structure factor  $S_{ii}(k)$, denoted hereafter as
$S(k)$.

An XRTS $W(k)$ calculation  needs the electron densities
 at an ion, and
the $S(k)$  of the system. The NPA
approach~\cite{Dagens1,PDW1} decomposes the total  charge
density into a superposition of effective one-body charge densities  and
structure factors,  and provides a comprehensive scheme based on
DFT. A number of NPA models are described in the literature,
e.g., those using ion-sphere models and other  prescriptions.
These affect how the chemical potential   is treated,
 and how the bound and free
electrons are identified\cite{NPA-PDW,NPA-other}. We use the
NPA model of Perrot and Dharma-wardana~\cite{NPA-PDW,NPA2} which uses a large
``correlation sphere'' of radius $R_c \sim 10 r_{ws}$,  where
$r_{ws}=\{3/(4\pi\rho)\}^{1/3}$ is the ion Wigner-Seitz radius. The electron
chemical potential is that of  non-interacting electrons at the interacting
density $n_e$ at $T_e$, as required by DFT. This model  accurately predicts phonons
(i.e., meV accuracy) in $2T$ WDM systems~\cite{CPP-Harb}. Thus
even the dynamical $S_{ii}(\mathbf{k},\omega)$ can be predicted when  XRTS
data at meV accuracy become available.

\begin{figure}[t]
\includegraphics[width=\columnwidth]{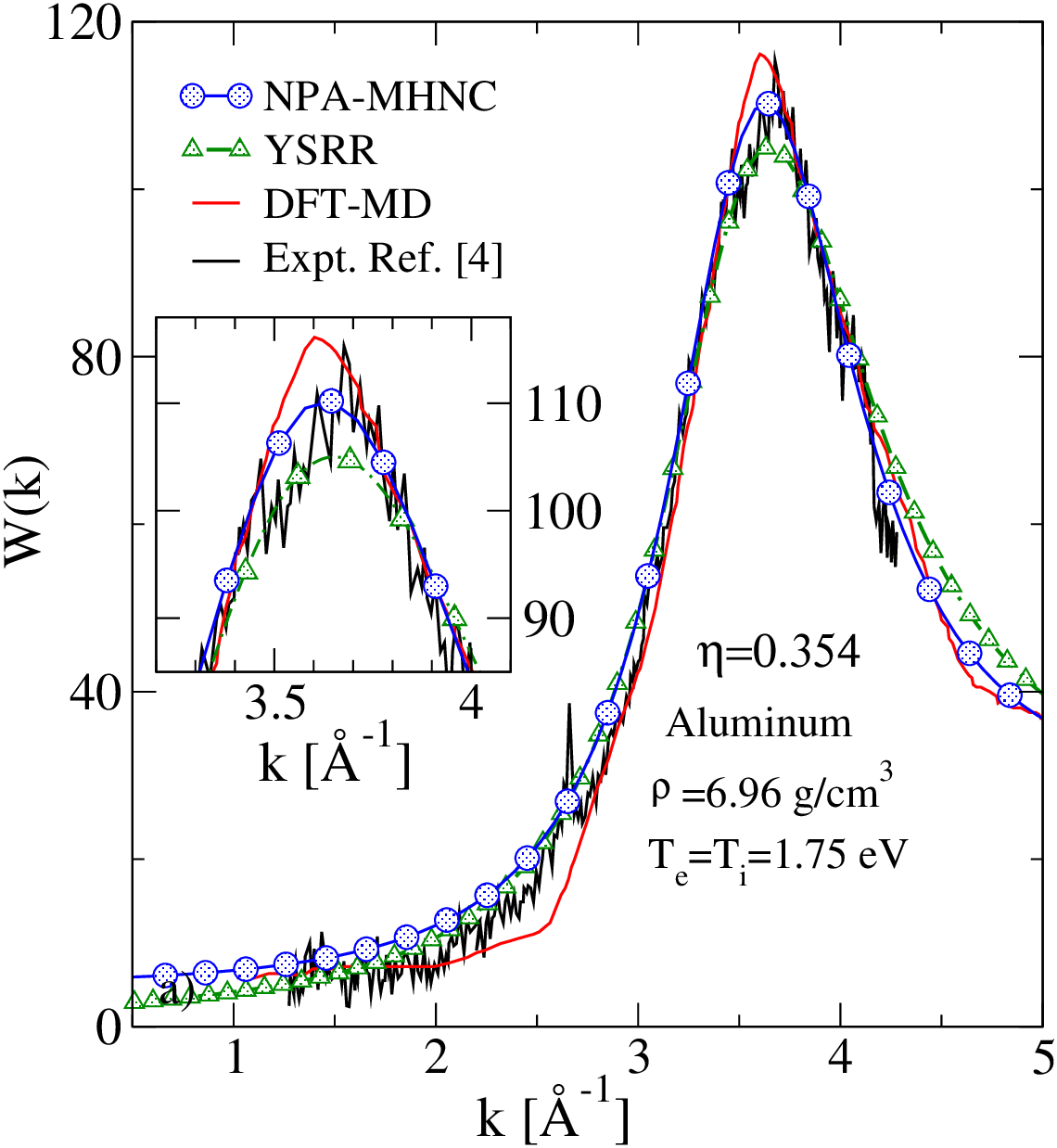}
\caption{The XRTS ion feature $W(k)$ of Ref.~\cite{Flet-Al-15}, and from DFT+MD,
  NPA-MHNC and YSRR models.}
\label{fig:WkFletcher}
\end{figure}

\begin{figure}[t]
\includegraphics[width=\columnwidth]{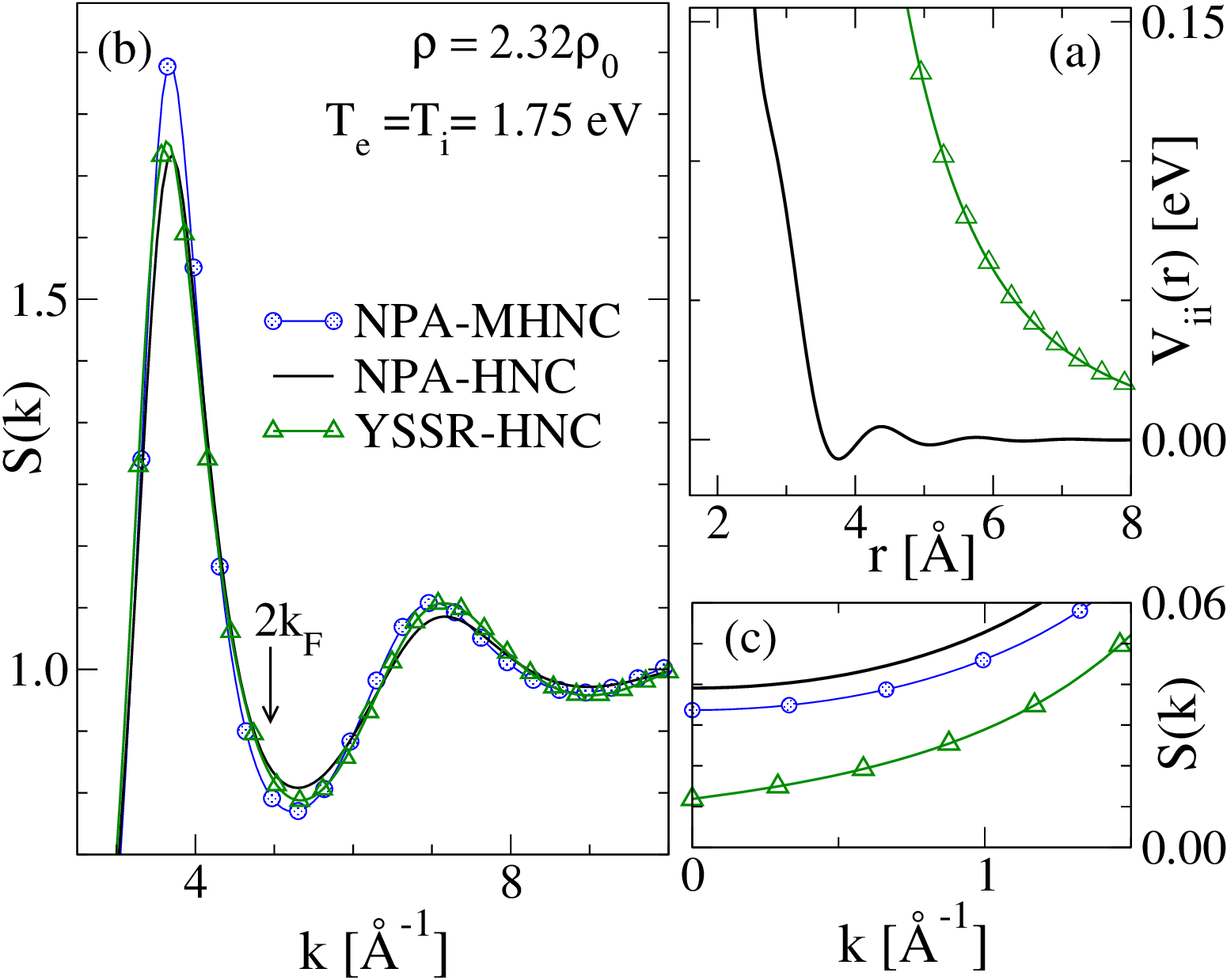}
\caption{ (a)NPA and YSRR
 potentials. (b) $S(k)$ from the $V_{ii}(r)$, using HNC and MHNC.
 (c) $k\to 0$ region of $S(k)$.}
\label{fig:vr_sk_fletcher}
\end{figure}

{\it Study of WDM-Al by Fletcher et al}$-$ Fletcher {\em et al}~\cite{Flet-Al-15} 
have studied compressed aluminum evolving across the melting line  into 
a WDM state, using XRTS. The data can be used to extract $S_{ii}(k)$,
the temperatures, 
compressibilities, pseudo-potentials etc., and  transport
 properties like  the conductivity. While DFT+MD can provide
 numerical results up to moderate $T$, such simulations fall short
 on clarifying the physics.
The physics comes out clearly in the
 DFT-NPA method~\cite{2Tpp},
 avoiding  misleading {\em ad hoc} YSRR-type models. 

The use of an equilibrium ($T_e=T_i$) model by
 Fletcher \textit{et al}  is justifiable for
 nano-second time sacles.
 The  NPA calculation provides the free-electron charge
 density $n_f(r)$ at an Al$^{3+}$ ion with
 $\mathcal{K}=\rho/\rho^0 = 2.32, \rho^0=$ 2.7 g/cm$^3$.
 The $n_f(r)$ is calculated using Kohn-Sham wave functions
 orthogonal to the  core states and hence core-valence Pauli
 effects and repulsions are
 correctly included. The corresponding
 electron-ion pseudopotential $U_{ei}(k) = n_f(k)/\chi_{ee}(k,T_e)$ uses
 the electron-electron  response function $\chi_{ee}$
  at  $T_e$. It uses a
 local-field correction (LFC).  The LFC uses the electron compressibility
 $\kappa_e$  matched
 to the finite-$T$  exchange-correlation functional~\cite{PDWXC}. 
 Given the pseudopotential
 $U_{ei}(k)$, the pair-potential  is evaluated as
 in Ref.~\cite{2Tpp}. Thus
  $V_{ii}(k)=4\pi Z^2/k^2 + |U_{ei}(k)|^2\chi_{ee}(k,T_e)$. 

The real-space form $V^\text{NPA}_{ii}(r)$
 can  be  compared with $V^{\text{ysrr}}_{ii}(r)$,
 where Fletcher {\it et al.}
use the Thomsas-Fermi 
 wavevector for Yukawa screening, with
 $k_s = k_{\text{TF}}$.
 The value of $\sigma$ is  4.9 a.u. correcting what may be
an error in  Ref.~\cite{Flet-Al-15} where $\sigma$= 9.4 a.u.
 is quoted~\cite{sigCorr}.
 The $W(k)$ from the  NPA and YSRR can be
compareed with the XRTS $W(k)$.
Fig.~\ref{fig:WkFletcher} shows $W(k)$ for Al, $\mathcal{K}=2.32$,
 at 1.75 eV, with the  $S(k)$ 
generated from a hyper-netted-chain (HNC) equation,
 i.e., without a bridge term. The Gibbs-Bogoluibov-Lado 
{\em et al.} (GBL) criterion~\cite{GBL, chenlai}
for the bridge function $B(\eta,r)$ gives a hard-sphere packing fraction $\eta=0.354$
for the modified-HNC (MHNC). The ion feature $W(k)$ from NPA-MHNC is
 in better agreement with experiment (Fig.~\ref{fig:WkFletcher}). 
%
%

The YSRR model was justified in Fletcher \textit{et al.} and in
 Wunsch \textit{et al}~\cite{Wunsch09}
for  inverting a given  $S(k)$ obtained from an
 MD simulation to extract
a  $V_{ii}(r)$ containing core-repulsion effects. 
The NPA calculation for Al at $T_e=1.75$ eV, $\mathcal{K}$=2.32
shows that the mean radius of the $n=2,l=0$ shell in Al, reflecting
the radius of the bound core  is 0.3552 \AA, while the YSRR potential reaches
 large values already by 2\AA.
We find that both the short-range part $\{\sigma/r\}^4$ and the
long-range part $\exp(-k_sr)/r$ of the YSRR form
are untenable.

The liquid-metal  community of the 1980s found 
that the inverse problem of extracting a potential from
the  $S(k)$ given in a limited $k$-range,  obtained from
DFT+MD or from experiment is misleading and  not
unique~\cite{March,AersCDW}. However, a parametrized  
 physically valid model (e.g.,the model of a pair-potential $V_{ii}(r)$
constrained via an
 atomic pseudopotential) together with a good $B(\eta,r)$~\cite{chenlai}
can successfully invert the MD-data.
However, the  DFT+MD step is quite unnecessary in most cases since
the $V^{\text{NPA}}_{ii}(r)$  and the $S(k)$ that
provide the  physics are easily evaluated from an NPA calculation.

 Aluminum at $\mathcal{K}=2.32$,
 $T_e=1.75$ eV, i.e., $T_e/E_F=0.085 $,  has a  near-degenerate
 electron gas with $V^{\text{NPA}}_{ii}(r)$  displaying Friedel oscillations,
 c.f., 
Fig.~\ref{fig:vr_sk_fletcher}(b).
The $S(k)$ from the NPA-HNC, NPA-MHNC  and YSRR-HNC 
are shown in
 Fig.~\ref{fig:vr_sk_fletcher}(a). The NPA-HNC $S(k)$ is
very similar to the YSRR-$S(k)$  but differs in the $k\to 0$ region, panel (c),
  and near 2$k_F$. The MHNC gives a higher main peak and an improved fit to the
experiment.  The YSRR
$S(k)$ grossly contravenes the compressibility sum rule, and
 a $B(\eta,r)$  would make matters worse.
Hence the YSRR-$S(k)$ is only from the HNC, withpout a $B(\eta,r)$.
The sum rule  $S_{ii}(0)=\rho T_i \kappa$  gives the compressibility
 $\kappa$  as 9.6 a.u.  from the YSRR $S(0)$,
while the NPA-HNC gives 29 a.u. and MHNC gives 26 a.u.
The NPA $\kappa$ is close to the 
the compressibility of $\sim$ 30 a.u. from an ABINIT calculation
(
 Thus the YSRR model is not trustworthy enough for EOS
 properties like the compressibility.

\begin{figure}[t]
\begin{minipage}[t]{0.49\columnwidth}
\includegraphics[width=\textwidth]{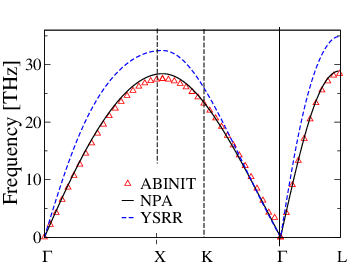}
\end{minipage}
\begin{minipage}[t]{0.467\columnwidth}
\includegraphics[trim=0.3cm 0 0 0, clip=true, width=\textwidth]{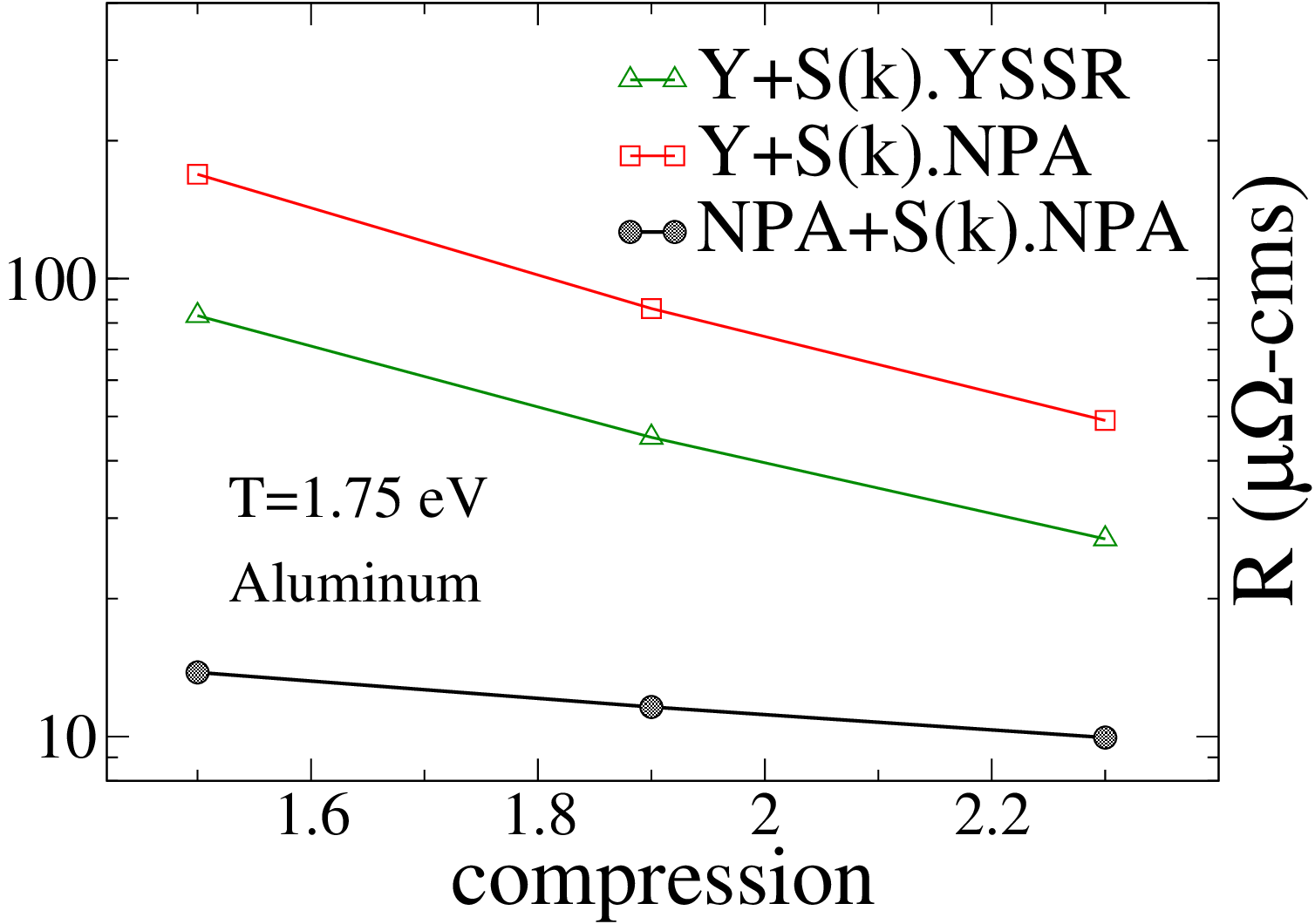}
\end{minipage}
\caption{(a)Longitudinal phonons of Al ($T_e$=1.75 eV, $\mathcal{K}=2.32$)
 from NPA, YSRR,  and ABINIT.
(b)Resisitivity $R$ from Yukawa and NPA $U_{ei}(k)$
and YSSR and NPA $S(k)$. }
\label{fig:Phon-Cond-Fle}
\end{figure}

Current XRTS spectra do not resolve the structure in $S_{ii}(\omega)$. However,
this is  approximated by the longitudinal-phonon spectrum which survives in
the liquid state. In Fig.~\ref{fig:Phon-Cond-Fle}(a), we present the longitudinal
phonon branch for an Al-FCC lattice at $\rho/\rho^0$ = 2.32 and $T_e$=1.75
eV, calculated from the NPA potential, the YSRR potential, and from an ABINIT
simulation. The unphysical ``stiffness'' of the YSRR potential leads to high 
phonon frequencies and a sound velocity more than  $\sim$20\% greater than the
NPA and ABINIT predictions.

We test  the validity of the Yukawa component  in the YSRR model
and the validity (or not) of the YSRR-$S(k)$ by calculating
 the electrical resistivity $R$  of aluminum for
 1.5 $<\mathcal{K}<$ 2.32 at $T_e$=1.75 eV. The Yukawa
pair-potential  $Z^2 \exp(-k_s r)/r$ arises from  the Yukawa pseudopotential
$U^y_{ei}(q)=-Z/(q^2+k^2_s)$ screened by the $k\to 0$  RPA dielectric
 function, i.e.,
$\epsilon(q)=1+(k_s/q)^2$.  We use the Ziman formula given
in~\cite{eos95},~Eq.~(31), with (a) NPA $S_{ii}(q)$ and the NPA pseudopotential
$U_{ei}(q)$, (b) NPA $S_{ii}(q)$ with  $U^y_{ei}(q)$, and YSRR $S(q)$ with
 $U^y_{ei}(q)$, to calculate  $R$ shown in\ Fig.~\ref{fig:Phon-Cond-Fle}(b). 
 The NPA-$U_{ei}(q)$ used with the NPA-$S(k)$ in the Ziman formula is a
well-tested method for weak scatterers~\cite{eos95,benage99}. It shows little
change in $R$  with $\mathcal{K}$  in this highly degenerate, compressed
regime, unlike the YSRR model. These results show that both the SRR potential,
 as well as the point-ion linear-screening (Yukawa) model are untenable. An
easily computable model with the correct physics is the DFT-based
NPA presented here. It generates a parameter-free simple (local) pseudopotential
$U_{ei}(q)=n_f(q)/\chi(q)$.


The claim that the YSRR potential ``accounts for the additional repulsion
from overlapping bound-electron wavefunctions''~\cite{Flet-Al-15} is not
confirmed by the calculated atomic structure of Al$^3+$ in the plasma. The
core-core interaction can be explicitly calculated from the core-charge density
using the analysis given in Appendix B of Ref.~\cite{eos95}. It
is quite negligible at the compressions used in
Refs.~\cite{Flet-Al-15,Ma-Al-13}.

\begin{figure}[t]
\includegraphics[width=\columnwidth]{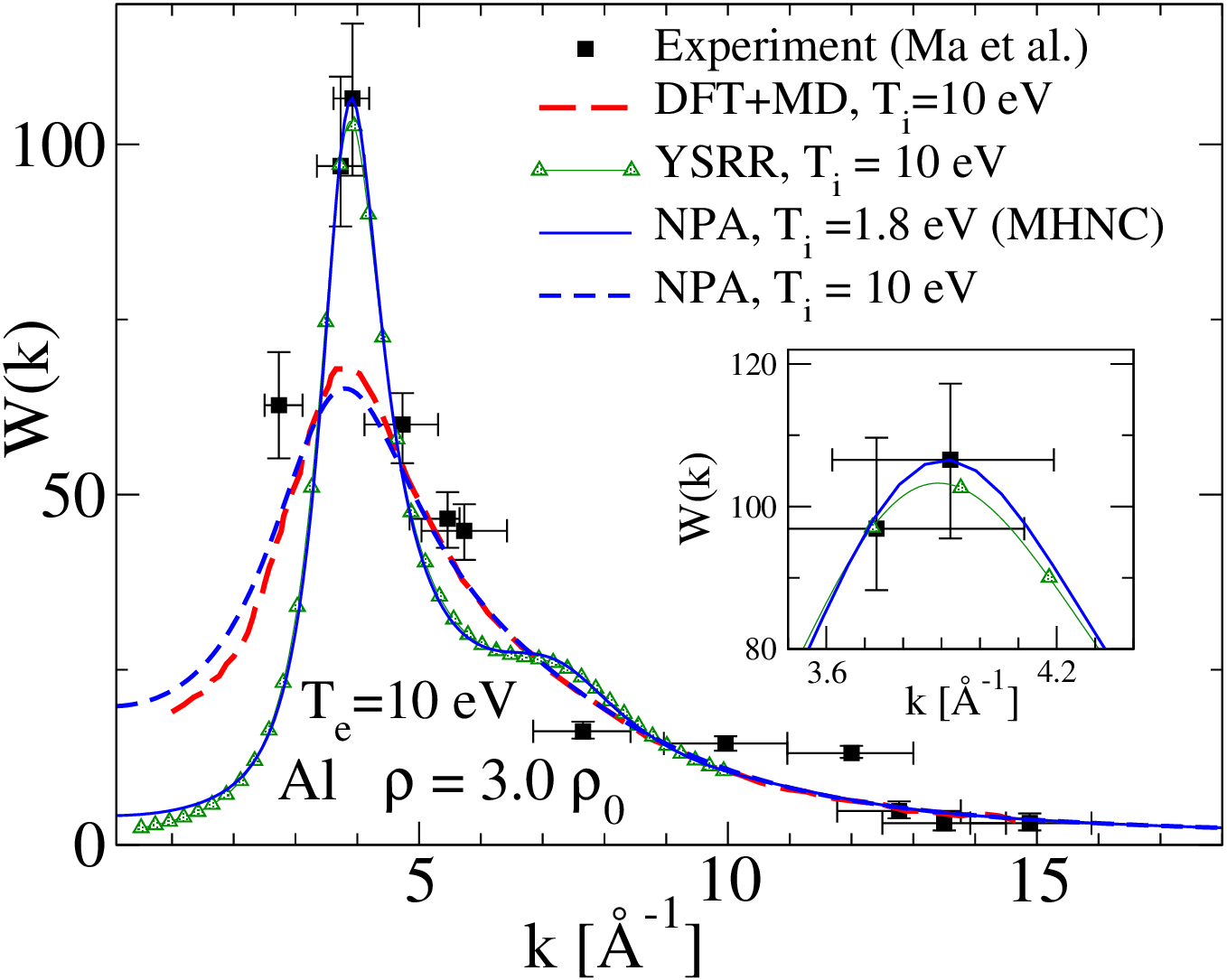}
\caption{XRTS ion feature $W(k)$ of Ma et al~\cite{Ma-Al-13} is compared with
 the $W(k)$ from YSRR, NPA, the DFT+MD $T_i=T_e$ calculation of
  R\"{u}tter \textit{et al.}, and from a 2$T$-NPA calculation
 with $T_i=1.8$ eV, $T_e=10$ eV.}
\label{fig:WkMa}
\end{figure}

{\it Study of WDM-Al by Ma et al}$-$ Ma \textit{et al}~\cite{Ma-Al-13}
studied  aluminum at  $\rho/\rho^0$ = 3 and  $T_e$=10 eV.
The YSRR model with $T_i=T_e$
misleadingly showed good agreement with the experiment.  
Ma \textit{et al.} state that the results from the YSRR model
demonstrate the ``importance of the short-range repulsion stemming from bound
electrons in addition to Yukawa-type linear screening caused by the free
electrons''. As already noted, the core-core repulsion term in aluminum
 is negligible~\cite{eos95}.

The ion feature at $T_e=T_i=10$ eV and $\mathcal{K}=3$ determined by the
DFT+MD  simulation of R\"{u}tter \textit{et al.}\cite{Ruther} 
disagreed with  the data of Ma \textit{et al.}~\ref{fig:WkMa}.  
Using additional DFT+MD simulations,
Cl\'{e}rouin {\em et al.}  proposed~\cite{Clerouin}  
a 2$T$ model where $T_i = 2$ eV while
$T_e = 10$ eV. Using the NPA-potential at $T_e$=10 eV, 
and  the MHNC with $\eta=0.367$ given by the GBL criterion,
we obtain an excellent fit to the Ma {\em et al.} data with $T_i =1.8$ eV,

The high-$k$ shoulders of  the $W(k)$ from the 2$T$ NPA-MHNC, and from
the YSRR are washed-out in the experiment, suggesting  more complexity
than in a 2$T$ system. The ion-subsystem  may be cold (at 1.8 eV), but
 containing an unknown high-$T$ component as well.

\begin{figure}[t]
\includegraphics[width=\columnwidth]{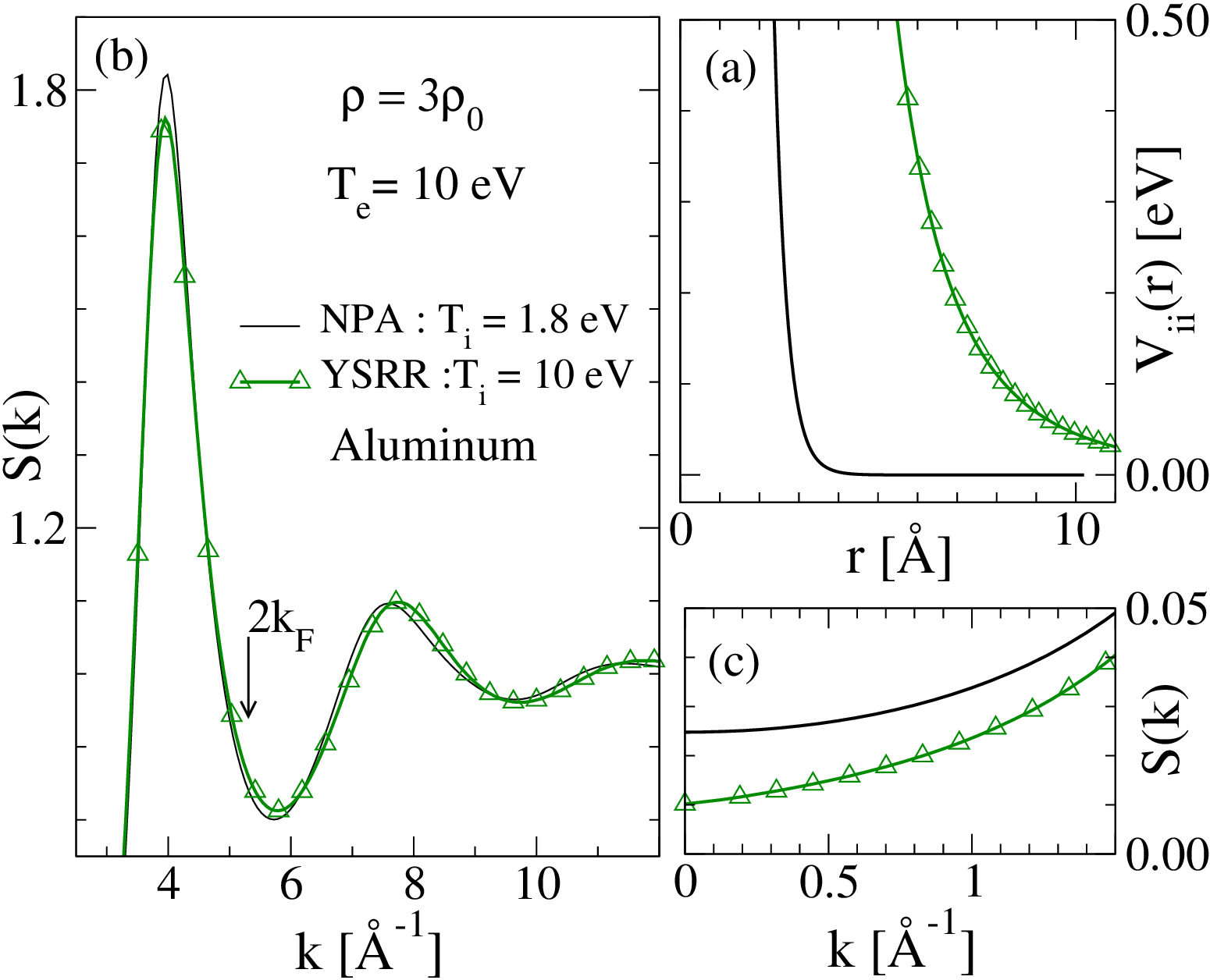}
\caption{(a) The NPA and YSRR  pair potentials.
 (b) the $S(k)$ from YSRR, and  NPA-MHNC (c)$S(k)$ for $k<1.5$.} 
\label{fig:vr_sk_ma}
\end{figure}

In Fig.~\ref{fig:vr_sk_ma}, the NPA and YSRR $S(k)$, pair-potentials $V_{ii}(r)$
  and the $k\to 0$ limit are presented in  panels (a)-(c). There are no
Friedel oscilaltions in  $V^{\text{NPA}}_{ii}(r)$ as $T_e$ is nearly six times
 higher  that in Fletcher {\em et al.}.
 However, assuming that $S_{ii}(k=0)=\rho T_i \kappa$ even for 2$T$ systems,
 the YSRR model gives  $\kappa$ = 1.06 a.u.,
 i.e., much lower than in the NPA result($T_i=1.8$ eV, $T_e=10$ eV) of 14.0 a.u.,
 which is  close to the ABINIT result of 16.4 a.u.
for the $T_e=10$ eV system.

\textit{Conclusion}$-$ 
The NPA model uses the electron density to obtain a
 a parameter-free calculations of the pseudo-potentials,
 pair potentials, structure factors, transport coefficients, XRTS spectrum etc
 of a given material.  We have used the NPA method to (a) investigate
 a  popular  model of a Yukawa-screened short-range-repulsive
 potential  and shown
that both its short-ranged part, as well as its screening part lead
to misleading compressibilities, phonons, transport data and XRTS spectra. 
(b) expose pitfalls in inverting structure data to obtain effective potentials,
(c) to focus on 
non-equilibrium states in laser-generated WDMs,  and  (d) present
 results from rapid, accurate NPA calculations.

\textit{Acknowledgments}$-$ This work was supported by grants from the Natural
Sciences and Engineering Research Council of Canada (NSERC) and the Fonds de
Recherche du Qu\'{e}bec - Nature et Technologies (FRQ-NT). We are indebted to
Calcul Qu\'{e}bec and Calcul Canada for generous allocations of computer
resources.


\begin{thebibliography}{[1]}
\bibitem{sigCorr}
We have contacted the authors of Ref.~\cite{Flet-Al-15} regarding this. If
$\sigma=9.4$, then the classical coupling constant at $r_\text{ws}=1$ a.u.
would be over 7800 at all temperatures as this term is independent of $T$. The
HNC does not converge for $\sigma\sim 9$, although it may be treated using MD.
\end{thebibliography}
\end{document}